\pdfoutput = 1
\documentclass[10pt,letterpaper]{article}
\usepackage{opex3}
\usepackage{graphics}

\begin{document}

\title{Optically probing long-range spatial correlation and symmetry in complex biophotonic architectures on the transparent insect wings }

\author{Pramod Kumar$^{a\dagger}$, Danish Shamoon$^{a}$, Dhirendra P. Singh$^{a}$, Sudip Mandal$^{b}$, Kamal P. Singh$^{a*}$}
\address{$^{a}$Femtosecond Laser Laboratory, Department of Physical Sciences, Indian Institute of Science Education and Research (IISER) Mohali,  Manauli-140306, Punjab, India.}
\address{$^{b}$Department of Biological Sciences, Indian Institute of Science Education and Research (IISER) Mohali,  Manauli-140306, Punjab, India.}
\email{$^{a*}$ kpsingh@iisermohali.ac.in} 
\email{$^{a\dagger}$pramod@iisermohali.ac.in}



\begin{abstract}
We experimentally probe complex bio-photonic architecture of microstructures on the transparent insect wings by a simple, non-invasive, real time optical technique. 
A stable and reproducible far-field diffraction pattern in transmission was observed using collimated cw and broadband fs pulses.
A quantitative analysis of the observed diffraction pattern unveiled a new form of long-range semi-periodic order 
of the microstructures over $mm$ scale. These observations agree well with Fourier analysis of SEM images of the wing taken 
at various length scales. We propose a simple quantitative model based on optical diffraction by an array of non-overlapping
microstructures with minimal disorder which supports our experimental observations.
Two different applications of our techniques are demonstrated. First, by scanning the laser beam across the wing sample we observed a rotation of the original diffraction profile which gives direct signature of organizational symmetry of microstructures.  
Second, we report the first optical detection of reorganization in the photonic architecture on the Drosophila wings by various genetic mutations. 
These results have potentials for design and development of diffractive optical components for applications and identifying routes to genetic control of biomemetic devices. 
\end{abstract}

\ocis{(140.7260) Photonic crystal; (050.5298) Diffraction; (050.1940), Fourior optics; (070.0070), Nanophtonics and photonics crystal;(350.4238).} 


\section{Introduction}
Nature has developed a remarkable variety of photonic structures in various insect wings \cite{pet2003, bar2011, Laszlo2011}. 
The cooperation of structural heterogeneities (regularity and irregularity) \cite{Shuichi2005, Poya11} in these natural 
bio-photonic architectures at optical wavelength scale interact with light in a specific way to produce various optical 
effects such as reflection \cite{Hafiz2006}, interference \cite{yosh2008}, diffraction \cite{Feng2011} fluorescence \cite{Eloise2011}, 
iridescence \cite{step2009}, and polarization sensitivity \cite{shinya2007}. Compared to the equivalent man-made optical devices, the biophotonic structures 
often possess greater complexity and could outperform their functions in some cases \cite{bar2011, kesong2011}. 
Due to the presence of multiple length scales and diversity in their design the optical behavior of 
such arrangements is still not fully understood. 

While the optical effects in the non-transparent wings of the butterfly and beetles have been well studied \cite{Laszlo2011, Poya11, shinya2007, Shao2010}, 
the transparent wings of many insects (Drosophila, bees, dragonfly) have attracted much less attention. 
Previous experiments on thin transparent wings observed various interference colours under white
light illumination and quantified their transmission and reflection spectra \cite{Valerie2009, Ekaterina2011}. 
Recently, the transparent wings of firefly have been exploited in designing
and optimizing optical components such as anti-reflection elements in laser diodes \cite{annick2013}. 
High resolution techniques like scanning electron microscopy (SEM) and atomic force microscopy have resolved   
structural complexity of transparent wings (of dragonfly) at micro and nano-scale \cite{Hooper2006}.
These studies revealed that the wing surface is decorated with a large number of micro-structures \cite{Laszlo2011, Shao2010, Ekaterina2011, heeso2010}. These microstructures were known to provide anti-wetting, self-cleaning and
aerodynamic properties to the wing surface \cite{wanger1996, ghir1974, sane2003}.
Recently, modulation of friction and adhesion on these microstuctures was also observed \cite{ashima}. 
However, the long-range organizing principles and symmetry of the complex photonic architectue in transparent wings
remain unexplored. A knowledge of their organization would be fundamental to understand how these systems have been naturally optimized to coherently manipulate light for various functions \cite{Mathias2010, Shinya2004}.
Using high resolution images from SEM/TEM to extract structural order over entire wing surface is not feasible and may even be
misleading. Therefore, an efficient and quantitative approach to explore correlations in spatial architecture over 
the entire length scale of the wing is desired.

One attractive possibility is to exploit high sensitivity of optical method that combines 
right spectral properties with high spatial coherence such as transmitted diffraction pattern \cite{Feng2011, kumar2013}. 
Remarkably, our technique is very sensitive and efficient to extract \emph{in situ} structural organization and symmetry of microstructures in a single-shot manner. 
We demonstrate that variability in the diffraction pattern is directly correlated with 
arrangements in the spatial distribution of the scales that cover the wing membranes. This optical imaging technique opens new ways for the 
non-invasive study and classification of different forms of irregularity in structural patterns. The investigation of natural 
microstructures lead to the knowledge that can be applied to design biomemetic materials with designer photonic properties with 
reduced engineering effort and cost of their fabrication \cite{bar2011, Poya11}. 

The aim of this paper is to address the following questions: Is there any long range ($mm$ scale) order and organizational symmetry 
in the array of microstructures on the transparent insect wing? 
Can we exploit complex diffraction pattern to quantitatively unveil new features in the arrangement of microstructure array? 
Answering these questions are crucial to understand design principles and multi-functional role 
of transparent wings \cite{Hooper2006, Shinya2004, ashima}. 
The paper is organized as follows. In section $2$, we describe our experimental setup. 
In section $3$, we report experimental observations of the diffraction pattern using various lasers. We also give 
a theoretical interpretation of these results using SEM analysis of the samples. In section $4$ and section $5$ we 
demonstrate applications of the optical techniques by measuring correlated diffraction pattern at various 
scales in the wing and quantifying the role of genetic mutations on the photonic architecture of the transparent wings of the Drosophila.

\section{Experimental set-up}\label{section-setup}
\begin{figure}[h]
\centering
\includegraphics[width=0.8 \textwidth]{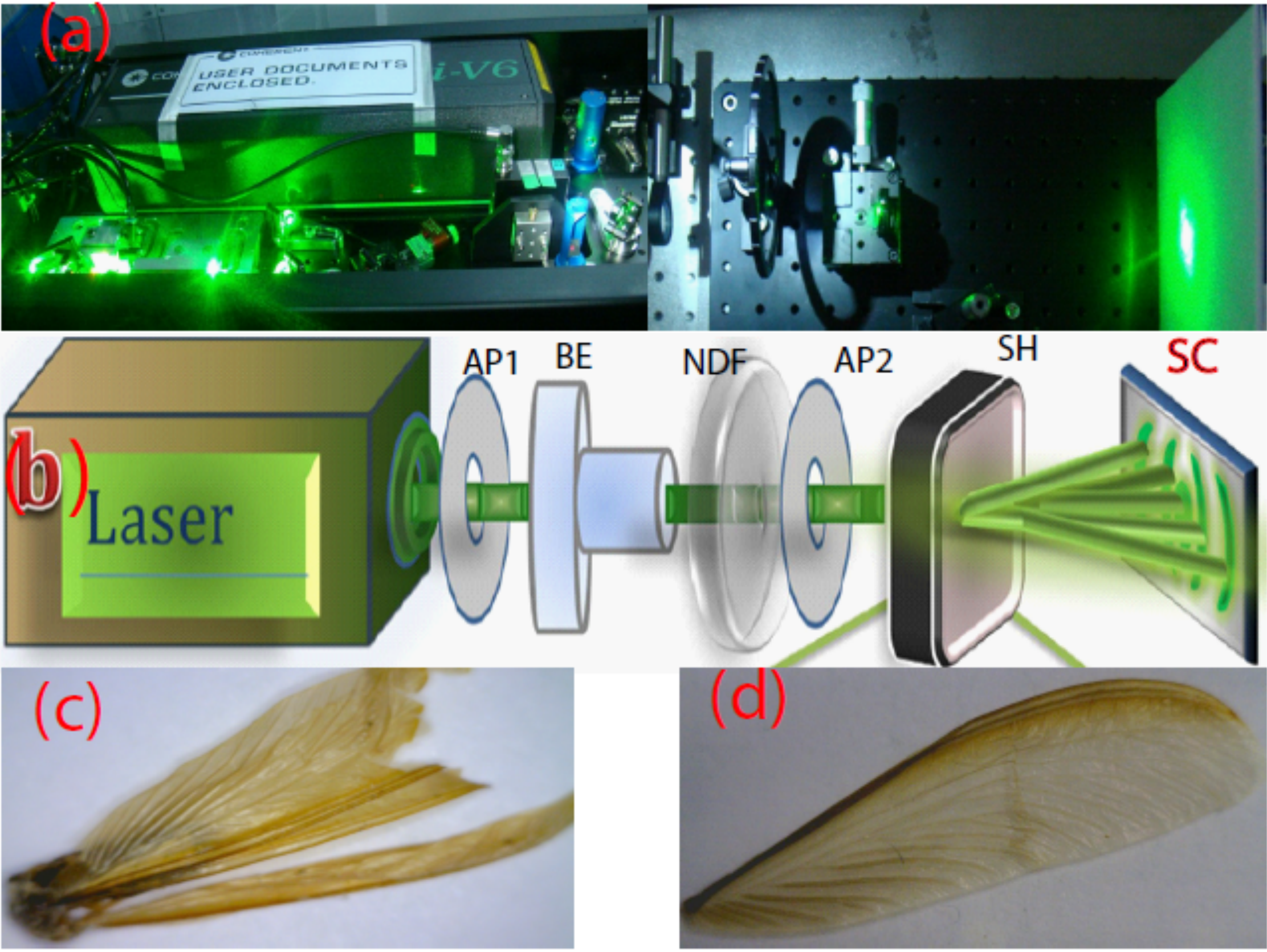}
\caption{ 
Experimental set-up and wing samples. (a) A picture of the femtosecond oscillator with the laser diffraction set-up for wing samples,
(b) schematic of the set-up. Various components are labeled as, SH: wing sample holder, AP1, AP2: irises, BE: (optional) beam expander, SC: screen, and NDF: neutral density filter. Wing to screen distance is $D=20.5~cm$.
(c)-(d) Pictures of the wing samples from the rain-fly.
}\label{fig1}

\end{figure}

A schematic diagram of our experimental set-up and its actual picture is shown in  Fig. \ref{fig1}. In our set-up a collimated laser beam passes 
through a wing sample that is mounted on a $xyz$ micrometer translation stage. The transmitted laser intensity was captured
through the wings by a digital camera and analyzed. 

We used both monochromatic cw lasers at two visible wavelengths in red and green ($\lambda=532~nm$ \& $632~nm$)
and femtosecond pulses centered in near IR range $800~nm$. These wavelengths are chosen to match the transparency window of the 
our wing sample. The typical intensity transmission coefficient of our insect wing is around $60 \%$ at these wavelengths. 
The $1/e^2$ full-waist of the collimated laser beams was around $1mm$ which is much smaller than the typical wing size $>1 cm$. 
The input beam profile of these lasers is shown in Fig. \ref{fig2}. The far-field diffraction pattern was captured on a white screen 
fixed at $D=20.5~cm$ from the wing sample. We have observed that the diffraction pattern is fully developed after few $cm$ from the wing 
and it simply diverges thereafter due to geometrical effect. It is worth 
mentioning that with this simple set-up, no preparation of the wing sample is required. In fact, 
it can be used for in vivo non-destructive imaging of the wing with the insect alive.
The laser powers were very low and no sign of optical damage was seen on the wing surface. 
To validate the sensitivity of our optical technique we also performed SEM images of the wing surfaces.   
In addition, we scanned the laser beam across the wing sample and the resulting laser diffraction was recorded in a single-shot 
manner to probe local variation in the wing structure. We also demonstrate how our optical technique can detect structural reorganization of microstructure array in genetically mutated wings of the Drosophila.   

\section{Results and discussion}\label{section-results}

\subsection{Observation and analysis of complex diffraction pattern}\label{subsection-diffraction}

\begin{figure}[h]
\centering
\includegraphics[width=0.8\textwidth]{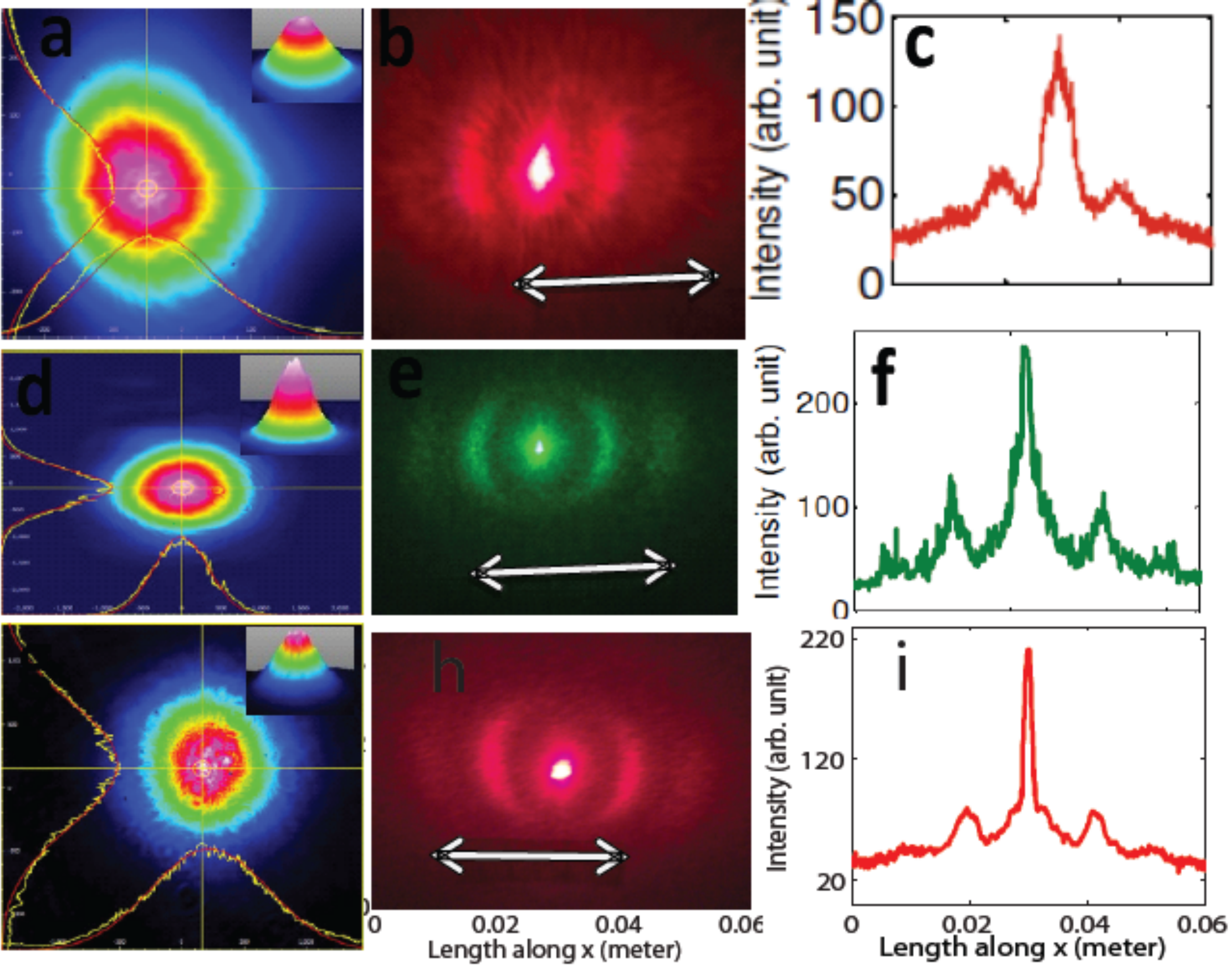}
\caption{ 
Experimental measurement of complex diffraction pattern. Left column (a, d, g) incident beam profiles for pulsed fs ($800nm$, 10 fs, 2nJ@78MHz) laser, cw $\lambda=532nm$ green solid-state 
laser and a cw red diode laser ($\lambda=632nm$) from top to bottom, respectively. Middle column (b, e, h) diffraction pattern on a screen using beam profiles in the left. Scale bar is $3 cm$. Right column (c,f,i) intensity cuts along x axis for the corresponding diffraction pattern in the left. }\label{fig2}
\end{figure}

Remarkably, a collimated laser beam ($1/e^2$ full waist around $1mm$) formed a stable and characteristic diffraction pattern after 
passing through the transparent wing-sample. The laser powers are around $5-200 mW$ which is below their damage thresholds.
The far-field diffraction pattern was observed for
(i) broadband femtosecond laser pulses centered at $800nm$ (top row in Fig. \ref{fig2}), (ii) a cw $532~nm$ green laser, and
(iii) a cw $632~nm$ red laser (bottom row in Fig. \ref{fig2}). Note that the intensity profile of the diffraction pattern was recorded 
on a calibrated screen for all the cases. For nearly Gaussian input beams, the observed intensity pattern $I(x,y)$ exhibited a bright central spot and up to two distinct higher order maxima in the form of curved lobes. These lobes are 
symmetrically located on both sides of the bright central spot. The femtosecond pulse is used to show the robustness of the diffraction pattern under broadband coherent source in the IR range. We have computed the corresponding spatial frequencies along x and y axes,
$k_x=(2\pi/{\lambda D}) x $ and $k_y=(2\pi/{\lambda D}) y$, respectively. Here D is the screen to wing distance and x, y are distances measured on the screen from the central spot. The position of the first lobe in the case of cw lasers correspond to spatial frequency of around $0.5 \times 10^{-6} radian/meter$. These spatial frequencies agree well with our theoretical analysis, as shown later. The corresponding intensity-cuts of the diffraction patterns along x axis (right columns in Fig. \ref{fig2}) also confirm
these values.

\begin{figure}[h]
\centering
\includegraphics[width=0.8 \textwidth]{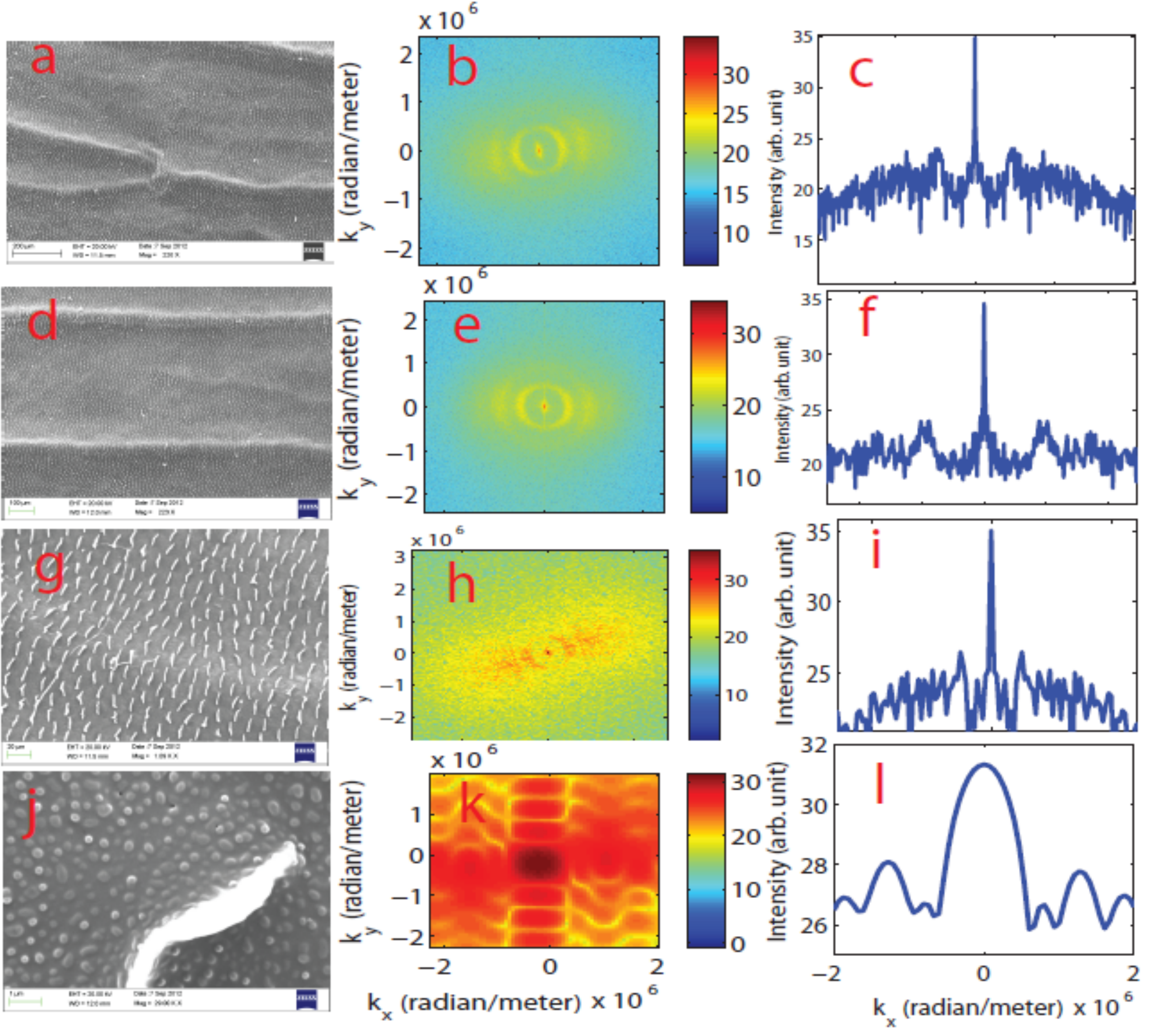}
\caption{ Left column: SEM images of the wing surface ($25nm$ gold coated) at varying length scales (a) $200 \mu m$, (d) $100 \mu m$, (g) $10 \mu m$, (j) $1 \mu m$. Middle column: computed FFT pattern shown as b, e, h, k for the corresponding images on its left. Right column: computed intensity cut for the corresponding FFT image. 
 }\label{fig4}
\end{figure}

\begin{figure}[h]
\centering
\includegraphics[width=0.7\textwidth]{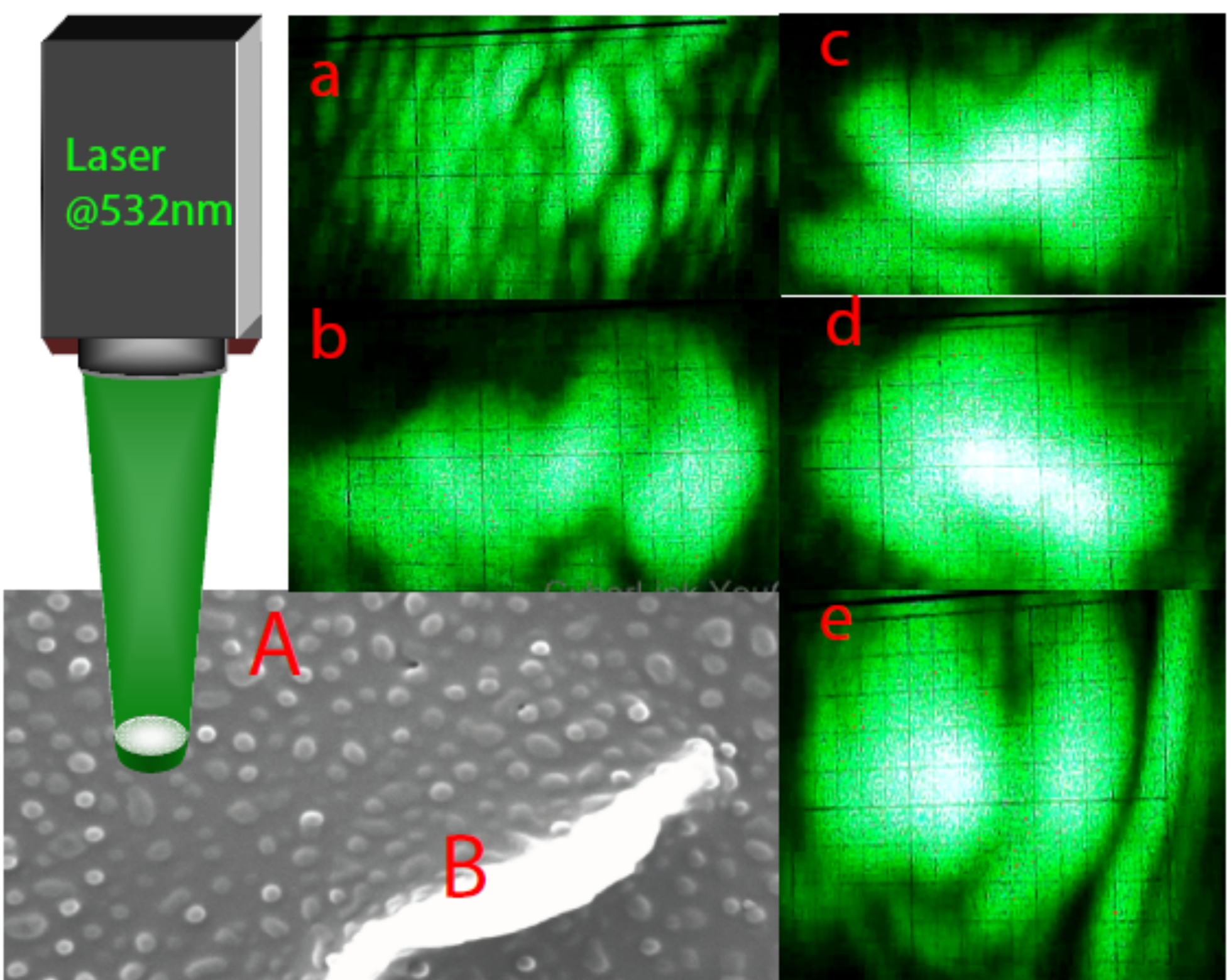}
\caption{ Laser diffraction from a single microstructure and background. (a) Schematic of the set-up. A triplet lens objective
was used to create $<5 \mu m$ spot diameter with the green laser beam. The wing was mounted on a micrometer translation stage.
(a-e) Various diffraction patterns of the background and microstructure when the micro-spot was scanned along the long axis of the wing. The pictures were taken on a calibrated screen at $5~cm$ away. 
}\label{fig3}
\end{figure}

To understand the formation of the diffraction pattern we performed SEM images of the wing surface (Fig. \ref{fig4}). 
The high resolution SEM images showed that the wing surface is decorated with a large number of non overlapping 
microstructures. The flat background of the wing exhibits nano-grain like features which are typically smaller than the wavelength of the light. These microstructures were elongated having typical length and width around $5-8 \mu m$ and $1-2 \mu m$, respectively.
Note that their dimensions are comparable to the used laser wavelengths, these act as efficient ``photonic elements''
that diffract light.  

To demonstrate how the observed far-field diffraction pattern emerges by a large number of microstructures (typically $10^{4-6}$ in beam waist of $1.0~mm$ ) we computed Fast Fourier transforms (FFT) of the SEM images at various scales. 
The SEM images contained total $N \times M$ pixels and it was sampled with the spatial resolution $\Delta x= L/M $ and $\Delta y= L/N$ along x and y axes, respectively. 
The complimentary Fourier domain then had spatial frequencies $k_{x}= m \Delta k_x$ and $k_{y}= n \Delta k_{y}$ where $m,n$ are 
integers and $\Delta k_{x,y}=2\pi/L$ determines resolution of the Fourier domain \cite{dav2003, mas2003}.
As we increased the area of SEM image  
to incorporate more number of microstructures (Fig. \ref{fig4}) the corresponding FFT showed an emergence of the higher order lobes.
The corresponding spatial frequency was around $0.5 \times 10^{-6} radian/meter$ that agrees quantitatively with the location of the experimental one.
One can define a quasi-periodic function $\Lambda(x, y)$ that determines the average spacing between microstructures.
As one can see in the SEM image that the quasi-period $\Lambda(x, y)$ is a function of position rather than a constant. Therefore, the input beam would be diffracted by an angle $\Lambda (x)\sin\Theta_{x}=m\lambda$  \cite{nan2013}.
The corresponding average period between the microstructures along the x-axis was around $12~\mu m$ 
that also agrees well with the SEM image analysis.

To further prove that the diffraction lobes are due to quasiperiodic organization of microstructure array, we performed experiment using tightly focused laser beam. Using a high numerical aperture (NA~0.2) triplet-lens objective we generated a micro-spot of full width $<5 \mu m$. Note that this spot size is comparable with the typical size of a single microstructure. We recorded the diffraction pattern on a screen kept about $5~cm$ away for two different cases when the laser spot is on the microstructrue and when the laser spot is on the background, i.e., between the two microstructures. We observed that these diffraction patterns are qualitatively different compared to the case when we used $1mm$ collimated beam. The FFT of high resolution SEM image with single microstructure also produced a similar complex diffraction pattern that matched with the experimental observations (see Fig. 3 and 4). 
This demonstrated that the observed diffraction is a result of quasi-periodicity in the array of microstructures. In the following we develop a theoretical model to explain our experimental results.

\subsection{Theoretical understanding of the experimental results}\label{subsection-theory}

We present a simple quantitative model to provide further insight into the experiment by generating 
the far-field diffraction from a two-dimensional array of non-overlapping microstructures. 
Each micro-structure was modeled by an amplitude transmission function $t(x,y)$. 
The total transmission $T$ of the wing was then due to $N \times M $ 
total number of microstructures arranged in a lattice with average distance between them being $d$.
Although, a similar jitter model was previously proposed to explain spectral properties of
an array of identical nonoverlapping grains \cite{mah1992,hen1978, mar1979, ru1993}. However, 
we introduce multiple disorders to model the laser diffraction through complex wing surface. 

We generated a two-dimensional total brightness distribution T(x,y) by creating square cells,
each containing one microstructure, arranged in a rectangular array of size $N \times M$. 
The distance between two microstructures in the square lattice was $d$ along the two orthogonal directions.
In the 
case of perfectly ordered array of identical microstructure, one can write, 
\begin{equation}
T(x, y) = t(x,y) \star \sum_{n=0}^{N-1} \sum_{m=0}^{M-1} \ \delta(x -n d) \delta (y - m d),
\end{equation}
where star denotes convolution and $\delta(x)$ is a Dirac delta function centered at $x=0$. 

To model structural organization of microstructures on the wing surface, we added disorder
in the shape of each microstruture, their position and orientation. The resulting stochastic
two-dimensional brightness function is given by,

\begin{equation}
T(x, y) = \sum_{n=0}^{N-1} \sum_{m=0}^{M-1} \ t(R - \alpha_{mn} - \xi(n) - \eta_{m,n} ),
\end{equation}

where $\alpha_{m,n}$ is the position vector of each microstructure, $\xi$ defines a two component 
stochastic vector quantifying the deviation of their position from their nominal center and $\eta_{m,n}$ is a set of variables
determining the shape distortions. The stochastic variables are delta correlated in position as,
\begin{equation}
 \left\langle \xi_{x,y} (n) \xi_{x,y}(n') \right\rangle  = A_{x,y} \delta(n-n'),
\end{equation}
where $A_{x,y}$ denote the amplitude of the noise along x and y directions. 

The Fourier transform of the transmission is given by $T(k_x, k_y) = FT(T(x,y))$ where $k_{x,y}$ denote
spatial frequencies \cite{yama2012}. The average normalized far-field diffraction pattern is given by,
\begin{equation}
I(x, y) = C \left| T(k_x,k_y)  \right|^{2} 
\end{equation} 
where $C$ is a proportionality constant. 
Note that the above equation is usually used for monochromatic incident field and it can be extended to a broadband pulse 
of width $\Delta\lambda$ that could lead to further detail of quasi-periodicity \cite{liu1997, yama2012}. 

\begin{figure}[h]
\centering
\includegraphics[width=0.8 \textwidth]{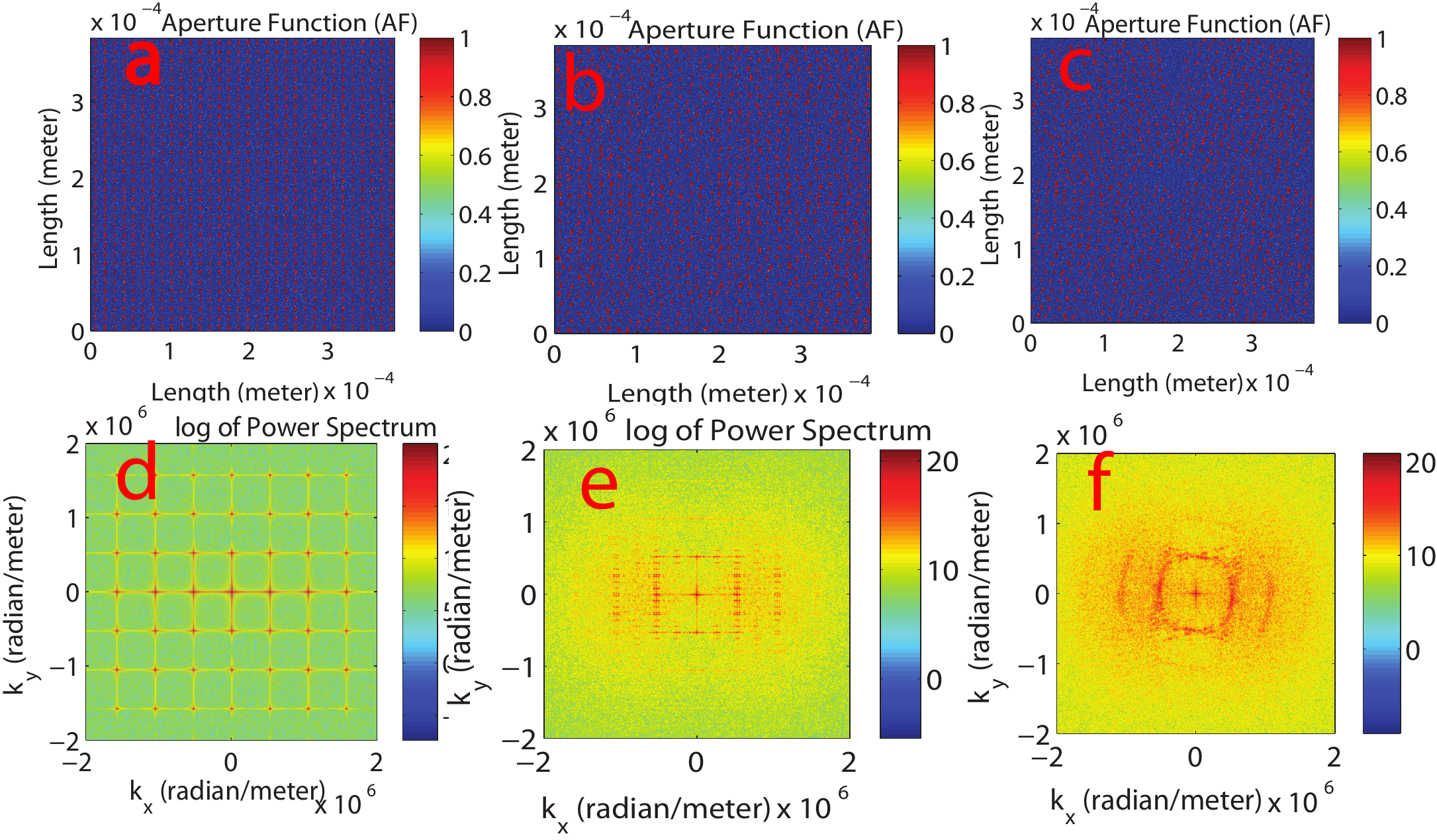}
\caption{Simulated 2d aperture functions and corresponding diffraction patterns. (a) The case of an array of 
microstructures arranged in squre grid without any disorder with the corresponding diffraction pattern in (d). (b) The case of 
minimal noise in shape and orientation of the elements with its diffraction pattern in (e). (c) The disordered aperture function
by adding random rotations on six patches within $\pm 15^\circ$ with corresponding diffraction pattern in (f).  
 }\label{fig5}
\end{figure}

We simulated the above model in Matlab by generating the transmission function $t(x,y)$ in the form of a curved ellipse
of about $2~\mu m$ width and about $8~\mu m$ length that is comparable to the dimensions observed in the SEM image.
Each microstructure consisted of $16 \times 16$ pixels where each pixel defined $750~nm$ resolution.
The total number of microstructures in our simulations were more than $10^3$.   
We verified that in the case of a perfect square grid of identical microstructures (Eq.~1), 
the simulated diffraction pattern consisted of ordered spots typical of regular diffraction grating (Fig. \ref{fig5}a, d).

To model our experiment we generated complex aperture functions by adding various kinds of disorders in the shape, orientation and position of each element.
In Fig. \ref{fig5}b, we show the disordered array which is produced by adding noise on the shape and orientation of $t(x,y)$. 
The variance in the fluctuations in the length and width was $8 \pm 4.5~\mu m$ and $2 \pm 0.75~\mu m$, respectively and their orientation was allowed to fluctuate randomly by $\pm 30^\circ$. The chosen parameters lie in the range seen in the SEM images. The corresponding far-field diffraction pattern for this organization showed that
higher than second order spots vanished but their organization was on the square grid. 

Finally, to create the various symmetries in the form of rings we randomly selected several patches and added
a global rotation by less than $\pm 15 ^\circ$ in order to create a wavy pattern on the grid structure. 
In this case, various symmetries overlapped and produced ring-like pattern similar to the experiment (see Fig. \ref{fig5}c). Note that the location of the first order peak correspond to spatial frequency $0.5\times 10^{-6}$ $radian/m$ close to the ones seen in the experiment as well as in the Fourier analysis of the SEM images. Though the minimally stochastic model uses many approximations, it can reproduce key signatures of the experiment quite well.

\section{Rotations of the diffraction pattern at various length scales}\label{subsection-rotation}
We can probe local variations in the structural arrangement across the wing by simply scanning the laser beam.
This is possible since the $1mm$ beam spot was much smaller than the wing size $>1cm$. To demonstrate this,
 we recorded the diffraction pattern at various scales by scanning the laser spot along the wing length (Fig. 6).
A rotation of the original diffraction pattern was observed for both the pulsed and cw lasers. This behavior directly 
reflects the local symmetry and its spatial correlation along the entire wing surface. It should be mentioned 
that one can easily vary the spot size to further probe local and the global organizations in a single shot manner. 
The observed rotation in the diffraction pattern suggests a systematic rotation in the arrangement of the hooks without much change in their density and shape.  

\begin{figure}[h]
\centering
\includegraphics[width=0.7 \textwidth]{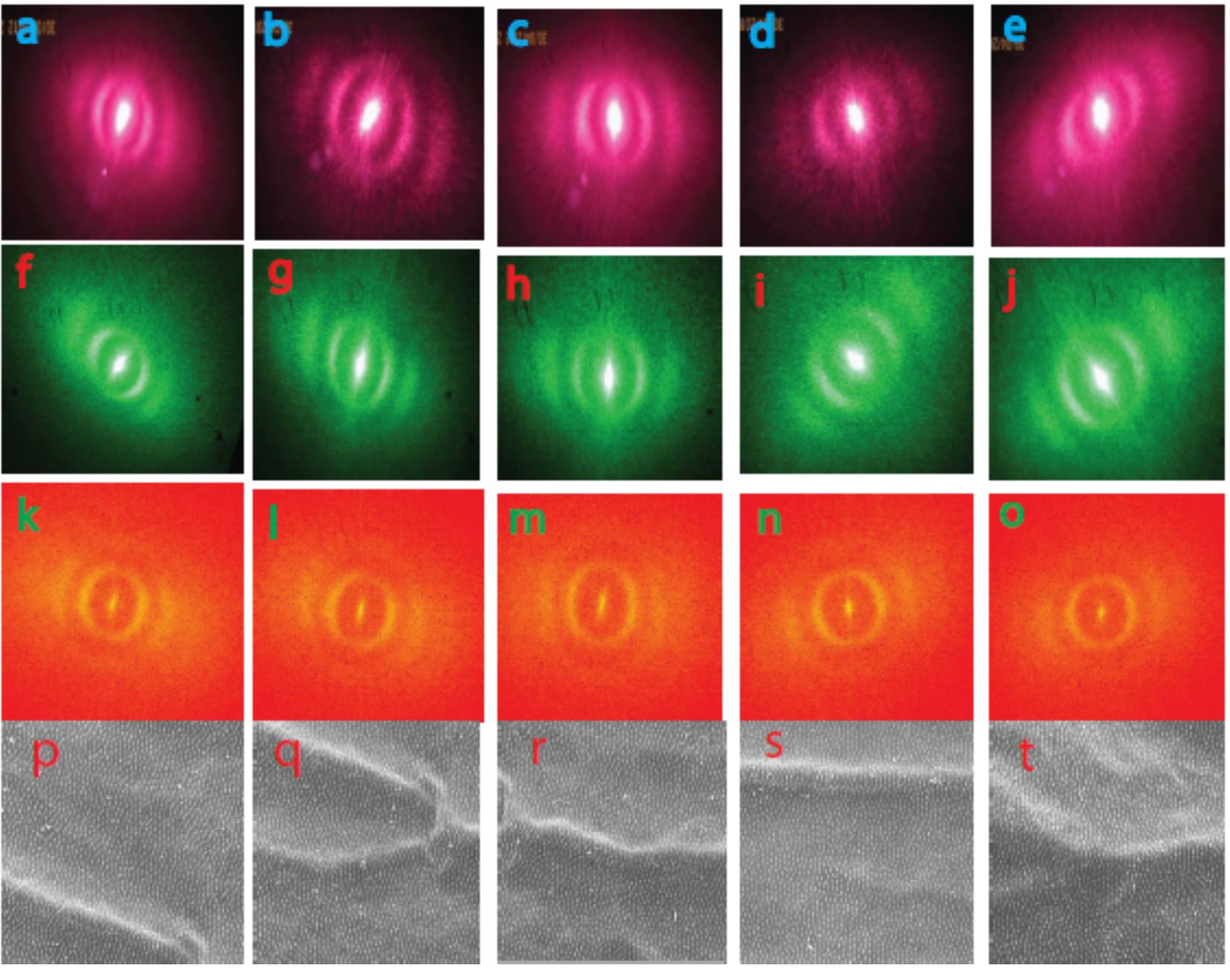}
\caption{ Rotation of the original diffraction profiles by scanning the laser beam across the wing sample. 
Top row (a-e) Experimental observations using broadband femtosecond laser, middle row (f-j): with a green cw laser.
(k-o): Theoretical rotation pattern produced by computing FFT of SEM images at different areas of the wing
shown below (p-t). 
 }\label{fig6}
\end{figure}

To verify this, we recorded SEM images of various portions of the wing and numerically computed FFT of the images (Fig. 6(k-o)). 
The SEM-FFT analysis also produced a similar rotation in the diffraction as shown in Fig. 6. This confirmed our 
optical observation. Therefore, our technique is very attractive and efficient to reveal the complex arrangements 
of millions of these photonic elements on the wing surface. We have also observed similar results in insect wings
of the Drosophila which suggest a generic nature of the reported phenomenon. 
The functional significance of these rotations and their development aspects requires further experimentations. 

\section{Genetic control of the photonic architecture }\label{section-application}
We report the first optical measurements of reorganization in mictrostructure array due to the genetic mutations 
on the transparent wings of the Drosophila melanogaster. The Drosophila wings are ideal system for this kind of study
since development, structure and function of the wings have been studied extensively and the roles of several genetic 
mutants are well characterized. For our analyses we selected two different mutants, Cyo and vg,  that either produces 
curly wings or generates small stumpy wing rudiments, respectively. The wild type unmutated wings were kept as controls (reference).
Although the developmental dynamics of Drosophila wings along with the implications of these mutants on the structural and functional 
aspect have been explored, much less is known about how these mutations affect organization of the microstructures on the wing surface. 
Using diffraction pattern in transmission offers us a unique optical technique to make a quantitative comparison among wings produced by various mutations. 
\begin{figure}[h]
\centering
\includegraphics[width=0.7 \textwidth]{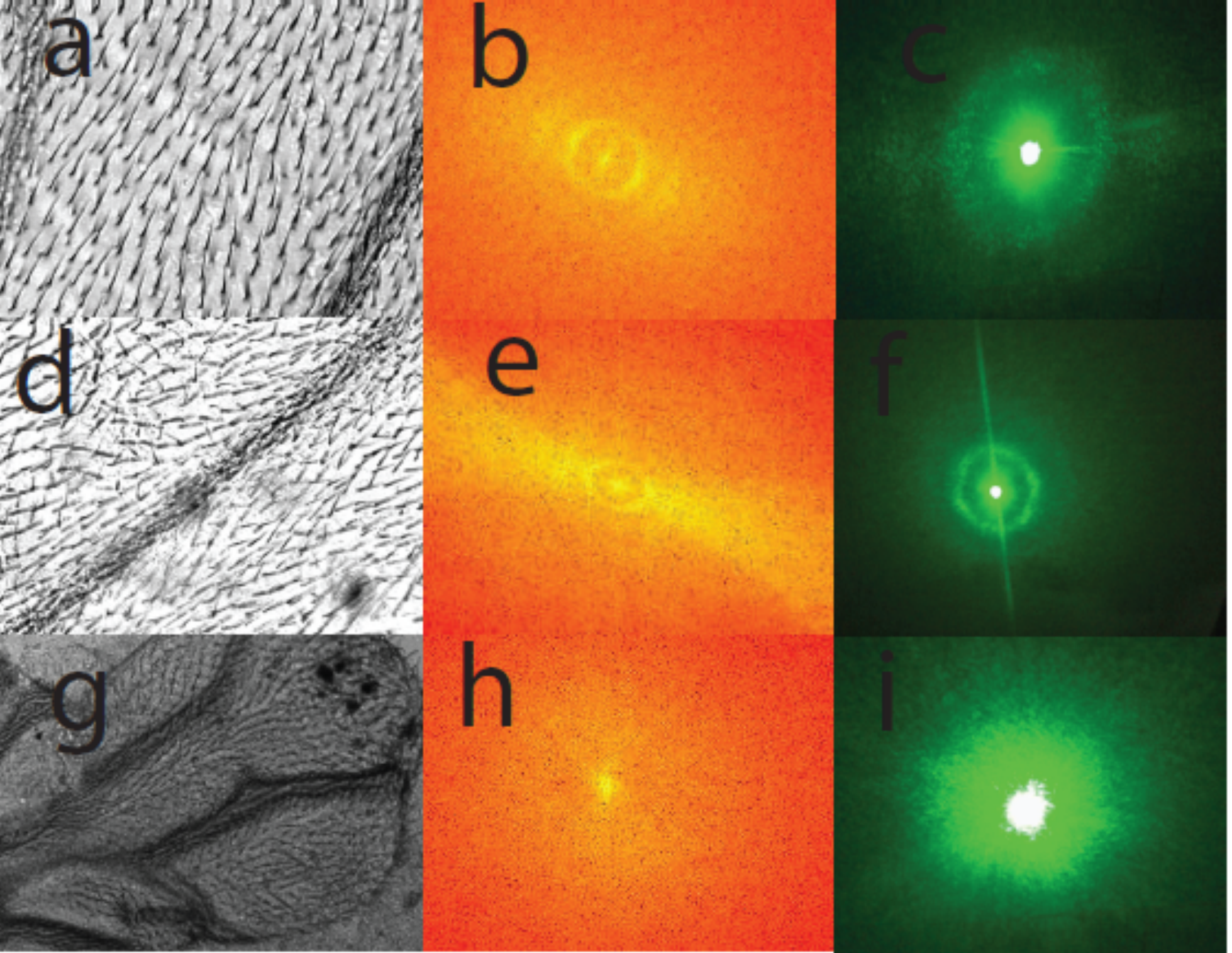}
\caption{ Reorganization of photonic miscrostructures by genetic mutations. Left column top to bottom shows the SEM images of the wing for wild-type unmutated wing, Cyo wing mutation, and vg mutation, respectively. Middle column: 
FFT of the SEM image for corresponding SEM image. Right column shows the observation of diffraction pattern with green laser for the corresponding mutations.}\label{fig7}
\end{figure}

First, we recorded the diffraction pattern from a normal fly-wing as our control. 
This clearly showed a ring shaped pattern revealing average periodicity of around $20~\mu m$. Compared to the control wing, in Cyo wings, the diffraction ring became smaller with a weak second order lobe appeared. 
This suggests that the average periodicity increased and their organization became more ordered compared to the normal wings.
In contrast, in vg wings the diffraction pattern was speckle-like without any higher order maxima and minima. 
This means that in this case both the symmetry and average periodicity is completely absent. Note that if such an 
information is attempted by SEM imaging, it would be very tedious and inefficient process. 
This is clearly illustrated in Fig. 7 where SEM images of various mutants are recorded and corresponding FFT is computed. 
The good matching between computational and experimental patterns again confirmed the implications of this technique.
 The spatial coherence of the laser and sensitivity of the diffraction pattern offer unique advantage over other methods.
 This experimental evidence of global correlation and its genetic control could be potentially useful 
to understand how one can manipulate genes to control the natural photonic architecture as well as for 
other potential biological applications in understanding structure-function relation in the genetic pathways.

\section{Conclusion and outlook}
In summary, we show that the diffraction pattern through the transparent insect wings is
correlated with the spatial organization of the microstructures at various length scales. 
We demonstrate that the microstructures on the transparent wings possess
a long-range quasi-periodic order and characteristic organizational symmetry as unveiled by an appearance of
the stable and robust diffraction pattern. These observations are in quantitative agreement 
with a Fourier analysis of high resolution SEM images of the wing surface. The existence of average periodicity
was also supported by observations of diffraction pattern of single microstructure and background using tightly focused laser beam.  Furthermore, we proposed a simple quantitative
model to explain our observations that showed the existence of minimal disorder in the microstructure organization. 
Two different applications of our optical technique were demonstrated. First, by scanning the laser beam across the wing 
surface, a rotation of the original diffraction pattern was observed that demonstrated symmetry in the spatial organization
of microstructures. Second, we reported first optical measurements on how various genetic mutations reorganize the biophotonic architecture.   

The proposed optical technique is potentially attractive to quantify natural photonic architecture on a large
variety of transparent insect wings in a single-shot manner. 
These tools would be crucial to understand design principles of photonic crystals with 
potential for biomimetic applications that may lead to novel optical devices \cite{bar2011, Mathias2010, gust2009}

\section{Acknowledgement}
We thank the DST and IISER Mohali for supporting this research. Pramod Kumar acknowledges postdoctoral fellowship from IISER Mohali. The invaluable help of Mrs. B. Basoya and Dr. Vinod Kumar are hereby grateful acknowledged. KPS acknowledges DST for financial support.
\end{document}